# *Operando* Label-free Optical Imaging of Solution-Phase Ion Transport and Electrochemistry

James K. Utterback,[1,2,†,*] Alex J. King,[2,3,4] Livia Belman-Wells,[2,5] David M. Larson,[2,‡] Leo M. Hamerlynck,[1] Adam Z. Weber,[2,3] Naomi S. Ginsberg[1,2,5,6,7,8,*]

[1] Department of Chemistry, University of California, Berkeley, California 94720, United States
[2] Liquid-Sunlight Alliance, Lawrence Berkeley National Laboratory, Berkeley, California 94720, United States
[3] Energy Conversion Group, Lawrence Berkeley National Laboratory, Berkeley, California 94720, United States
[4] Department of Chemical and Biomolecular Engineering, University of California, Berkeley, California 94720, United States
[5] Department of Physics, University of California Berkeley, Berkeley, California 94720, United States
[6] STROBE, National Science Foundation Science and Technology Center, University of California Berkeley, Berkeley, California 94720, United States
[7] Materials Science Division and Molecular Biophysics and Integrated Bioimaging Division, Lawrence Berkeley National Laboratory, Berkeley, California 94720, United States
[8] Kavli Energy NanoSciences Institute at Berkeley, Berkeley, California 94720, United States
[†] Present Address: Sorbonne Université, CNRS, Institut des NanoSciences de Paris, INSP, 75005 Paris, France
[‡] Present Address: Twelve, Berkeley, California 94710, United States

*Correspondence to: nsginsberg@berkeley.edu, james.utterback@insp.jussieu.fr

**ABSTRACT**

Ion transport is a fundamental process in many physical, chemical, and biological phenomena, and especially in electrochemical energy conversion and storage. Despite its immense importance, demonstrations of label-free, spatially and temporally resolved ion imaging in the solution phase under *operando* conditions are not widespread. Here we spatiotemporally map ion concentration gradient evolution in solution and yield ion transport parameters by refining interferometric reflection microscopy, obviating the need for absorptive or 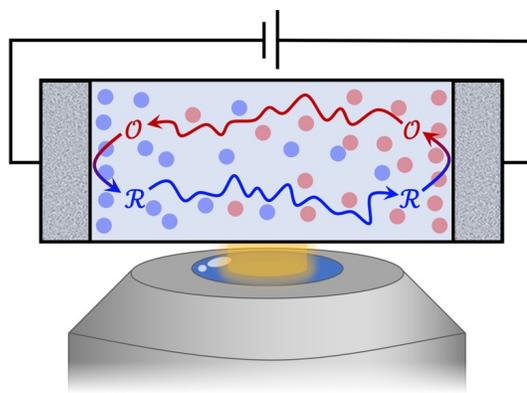 fluorescent labels. As an example, we use an electrochemical cell with planar electrodes to drive concentration gradients in a ferricyanide-based aqueous redox electrolyte, and we observe the lateral spatiotemporal evolution of ions via concentration-dependent changes to the refractive index. Analysis of an evolving spatiotemporal ion distribution directly yields the diffusivity of the redox-active species. The simplicity of this approach makes it amenable to probing local ion transport behavior in a wide range of electrochemical, bioelectronic, and electrophysiological systems.

Achieving a detailed picture of ion transport in the solution phase under (electro)chemically active conditions remains a major challenge that limits advances in applications ranging from solar fuel conversion and bipolar membranes to the understanding of electrophysiology.[1–4] Ion transport is a multiscale problem of diffusion and electrostatics that is dictated by nanoscale solvent structuring, mesoscale impacts of local environmental structure, and microscale transport.[5,6] Typically, ion transport within electrochemical systems is probed indirectly with bulk measurements of voltage and current, leaving the species concentration profiles and transport properties to be inferred through modeling and simulations.[7–9] There exist powerful techniques for probing electrode and electrolyte structure or steady-state ion distributions,[10–20] and absorptive and fluorescent labels as well as Raman-active modes have been used to probe ion transport with high resolution in specific cases,[9,21–32] but there is an outstanding need for direct universal measurements of ion transport in native microenvironments with spatiotemporal resolution. Optical detection of electrochemical processes via voltage-induced refractive index changes at electrodes also has a long history[11–13,33–36] and has more recently been used for imaging electrochemical processes at electrode–electrolyte interfaces,[35–39] electric double layer dynamics,[40] and ion intercalation in battery electrodes.[41] Nevertheless, dynamic *operando* measurements of ion transport in solution that do not depend on optical resonances, Raman-active modes, or ion proximity to an electrode surface or plasmonic structures remain difficult. Although they generally lack readily accessible optical signatures, ionic species do in general perturb the refractive index of their environments. The primary challenge of this contrast mechanism is the fact that solution-phase ionic species typically introduce a relatively small perturbation to the solution's refractive index.[42,43] For example, steady-state electrochemically induced ion gradients have been detected long ago but required prolonged integration times.[11–13] Therefore, to observe and extract the *dynamics of ion transport* requires detecting the small changes to a solution's refractive index as a function of position with sufficient time resolution. As a result, a universal *operando* approach for imaging ion dynamics in solution with sufficient spatiotemporal resolution has remained elusive.

Here we refine interference reflection microscopy[35,36,44,45] to spatiotemporally resolve electrochemically-induced ion concentration gradient evolution in solution with sub-mM, sub-µm and ms sensitivity without using additional labels or electrode enhancement effects. *Operando* experiments are achieved with an electrochemical cell compatible with optical microscopy such that the solution between parallel vertical electrodes can be observed under applied bias. Electrochemical reaction-induced concentration changes of an aqueous ferricyanide-based redox electrolyte cause measurable changes in the reflectance at the coverslip–solution interface down to ~$10^{-4}$ on a CMOS camera. Spatiotemporally resolving the evolution of the concentration gradients that develop between the electrodes via wide-field imaging directly yields the diffusivity of the redox-active ions in solution. Because measurements are made off-resonantly far from any optical absorption peaks, the observed signal is due to the real part of the refractive index, suggesting a universal method to spatiotemporally resolve the transport of ionic species. Multiphysics continuum simulations corroborate the experiments and analysis, showing consistency with the measured signal amplitude and spatiotemporal dynamics.

To spatiotemporally measure ion transport in solution we use interference reflection microscopy, which can detect small changes in the refractive index that occur due to voltage-induced changes in ion concentration. Wide-field imaging is achieved using an LED light source, a high numerical aperture oil-immersion objective, and a CMOS camera (Figure 1, see Supporting

Information (SI) for further details). The light intensity on the camera is proportional to the reflectance $R$ from the quartz–solution interface as a function of position and time. Thus, a normalized change in intensity in response to a change in solvated ion concentration is equal to the normalized change in reflectance, $\Delta R/R = (R - R_0)/R_0$, where $R_0$ is the reflectance before a perturbation is applied (i.e., applied voltage or static concentration change). To establish the method, we chose to use the model redox-active system of aqueous ferricyanide/ferrocyanide ($[Fe(CN)_6]^{3-}/[Fe(CN)_6]^{4-}$) with a supporting electrolyte of $K_2SO_4$—a very thoroughly documented system with well-understood reaction and transport properties.[6,46] Importantly, the light source is not resonant with the absorption feature of any species in solution (Figure S1) to demonstrate the generality of the approach. To generate ion concentration gradients and probe them *in situ*, we designed an electrochemical cell that features a microfluidic channel that has parallel Pt electrode walls (Figure 1, S2). The vertical electrodes extend 3 μm above the quartz–solution interface to the PDMS roof of the 2 mm-long channel; this geometry is designed to eliminate axial concentration gradients and render the mass transfer dynamics one-dimensional. Applying a voltage across the cell drives the reaction $[Fe(CN)_6]^{3-} + e^- \rightleftharpoons [Fe(CN)_6]^{4-}$ in the oxidative (reductive) direction at the anode (cathode) without changing the average concentrations throughout the system.

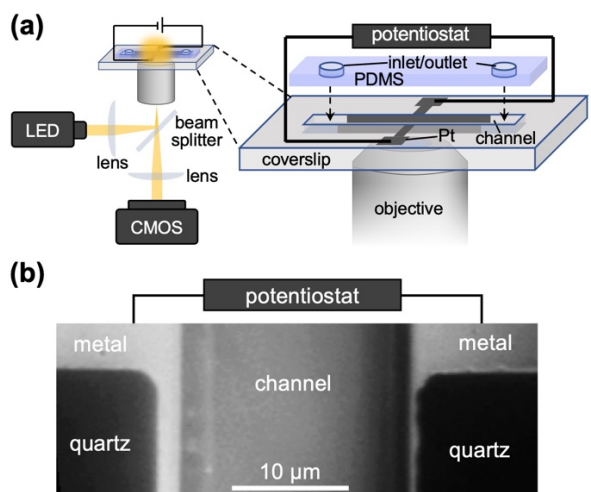

**Figure 1.** *Operando* interferometric reflection microscopy. (a) Measurement schematic and custom electrochemical cell. (b) Optical reflectance image of empty microfluidic electrochemical cell.

To calibrate the concentration-induced changes in reflectance, we measure $\Delta R/R$ of 595 nm light for static solutions on a coverslip as a function of ion concentration relative to $R_0$ for pure water. Droplets of solution containing a series of concentrations of dissolved salts are placed on a coverslip in air and we immediately measure the intensity on the camera while focusing on the coverslip–solution interface. $\Delta R/R$ decreases linearly with increasing concentration of $K_2SO_4$, $K_3[Fe(CN)_6]$, and $K_4[Fe(CN)_6]$—as well as when combined in equal parts—over the range measured (Figure 2a). This trend originates from the change in refractive index with solute concentration. At a wavelength of 595 nm, the absorbance of each species in solution is negligible (Figure S1), thus the reflectance is dominated by the real part of the refractive index:

$$R \approx (n_{\text{sub}} - n_{\text{soln}})^2/(n_{\text{sub}} + n_{\text{soln}})^2, \text{ (eq 1)}$$

where $n_{\text{sub}}$ and $n_{\text{soln}}$ are the real parts of the refractive index of the substrate (coverslip) and the solution, respectively. For relevant salt concentrations such as those used here,[5,6] the mole fractions are sufficiently small that solution refractive index is well known to change linearly with the solute concentration:[42,43]

$$n_{\text{soln}} = n_{\text{water}} + \sum_i K_i x_i, \text{ (eq 2)}$$

where $n_{\text{water}}$ is the refractive index of the neat solvent, $x_i = [\text{solute } i]/([\text{solvent}] + \sum_i [\text{solute } i])$ is solute $i$'s mole fraction, and $K_i$ is a coefficient that is known to depend on the polarizabilities of the solvated species and the volume displacement they induce on the solvent.[42,43] By assuming (i) that each solute approximately contributes independently,[43] (ii) that $\sum_i K_i x_i \ll n_{\text{sub}}, n_{\text{water}}, |n_{\text{sub}}^2 - n_{\text{water}}^2|$, and (iii) that $[\text{solute } i] \ll [\text{solvent}]$, eqs 1 and 2 reduce to first order in $\Delta[\text{solute } i]$ to an expression for the ion-induced imaging contrast,

$$\frac{\Delta R}{R} \approx -\frac{4 n_{\text{sub}}}{(n_{\text{sub}}^2 - n_{\text{water}}^2)[\text{solvent}]} \sum_i K_i \Delta[\text{solute } i], \text{ (eq 3)}$$

where the sum is taken over all solute species in solution. This expression fits the measured data in Figure 2a with high fidelity (see Table S1 for fitted values of $K_i$). Equation 3 indicates that the total contrast is determined by the weighted sum of the $\Delta[\text{solute } i]$ with weights $K_i$ while the nature of the substrate and solvent act as a prefactor that determines the sign and modulates the magnitude of the contrast. For salts in solution, $K_i$ is typically dominated by the anions because their polarizability is often larger than that of many cations.[13,43] Additionally, our simulations described below show that electrochemically-induced $\Delta[\text{solute } i]$ is large for the redox-active species and relatively small for spectator ions under our experimental conditions (Figure S3). These two factors result in $\sum K_i \Delta[\text{solute } i]$ being dominated by $[\text{Fe(CN)}_6]^{3-}$ and $[\text{Fe(CN)}_6]^{4-}$ in our measurements. Under these approximations, we use the data in Figure 2a and room temperature values of $n_{\text{sub}}$, $n_{\text{water}}$ and [water] to estimate that $K_{\text{K}_4[\text{Fe(CN)}_6]} - K_{\text{K}_3[\text{Fe(CN)}_6]} \approx 1.8$ and find that, for equal and opposite changes in the concentrations of $[\text{Fe(CN)}_6]^{3-}$ and $[\text{Fe(CN)}_6]^{4-}$ in an electrochemical cell, eq 3 reduces to $\Delta R/R \approx -\alpha \Delta[[\text{Fe(CN)}_6]^{4-}] = +\alpha \Delta[[\text{Fe(CN)}_6]^{3-}]$, with $\alpha = \frac{4 n_{\text{sub}}}{(n_{\text{sub}}^2 - n_{\text{water}}^2)[\text{water}]} \left( K_{\text{K}_4[\text{Fe(CN)}_6]} - K_{\text{K}_3[\text{Fe(CN)}_6]} \right) \approx 0.3$ M$^{-1}$. Assuming the illumination is sufficiently coherent, the 12-bit detector sensitivity limit yields $\Delta R/R \sim 10^{-4}$, which permits sub-mM sensitivity to concentration changes in the present case.

Interferometric reflection microscopy is sensitive to electrochemically-induced changes in ion concentration changes well away from the electrodes. The electrochemical cell was loaded with 100 mM $K_3[\text{Fe(CN)}_6]$, $K_4[\text{Fe(CN)}_6]$ and $K_2SO_4$ (Figure 2b). Below, $[\text{Fe(CN)}_6]^{3-}$ and $[\text{Fe(CN)}_6]^{4-}$ are referred to as $\mathcal{O}$ and $\mathcal{R}$, respectively. Experiments begin by collecting images at uniform concentration at equilibrium with no applied voltage and at open circuit, and >100 frames are averaged to serve as $R_0$ to calculate $\Delta R/R$. Upon applying a voltage of +0.4 V across the electrodes, $\Delta R/R$ increases near the anode and decreases near the cathode (Figure 2b). This behavior is expected for our system given the fact that $K_\mathcal{R} > K_\mathcal{O}$ (see eq 3, Figure 2a, S4), thus $\Delta R/R$ decreases (increases) near the cathode (anode) where $\mathcal{O}$ ($\mathcal{R}$) reacts to produce $\mathcal{R}$ ($\mathcal{O}$). One-dimensional $\Delta R/R$ profiles at 5 different times following the application of the voltage (Figure 2c), obtained by averaging images along the channel axis, show clear symmetric behavior of the gradient that develops between the electrodes. As seen in time traces collected with 1.5 ms time resolution for selected positions between the electrodes (Figure 2e, circles), the system reaches steady state within ~100 ms of the voltage being applied and relaxes back to the equilibrium state when the voltage is released. The corresponding current transient appears in Figure S5. Applying an alternating square wave of ±0.4 V peak-to-peak across the electrodes further confirms the assignment of the voltage-induced signal to changes in [$\mathcal{O}$] and [$\mathcal{R}$], as it demonstrates a reversible process as expected for the reaction $\mathcal{O} + e^- \rightleftharpoons \mathcal{R}$ (Figure S6). Such voltage-induced changes in $\Delta R/R$ are not observed when the cell is loaded with pure water or the supporting electrolyte $K_2SO_4$ alone (Figure S6). The diffraction-limited behavior near the electrodes is suggestive of electric double-layer formation or electrochemically-induced surface plasmon resonance changes, which have been thoroughly treated in previous reports.[34,36,37,40] Here we focus on contrast and transport in the mass transport boundary layer that in general lies between the electric double layer and bulk solution.

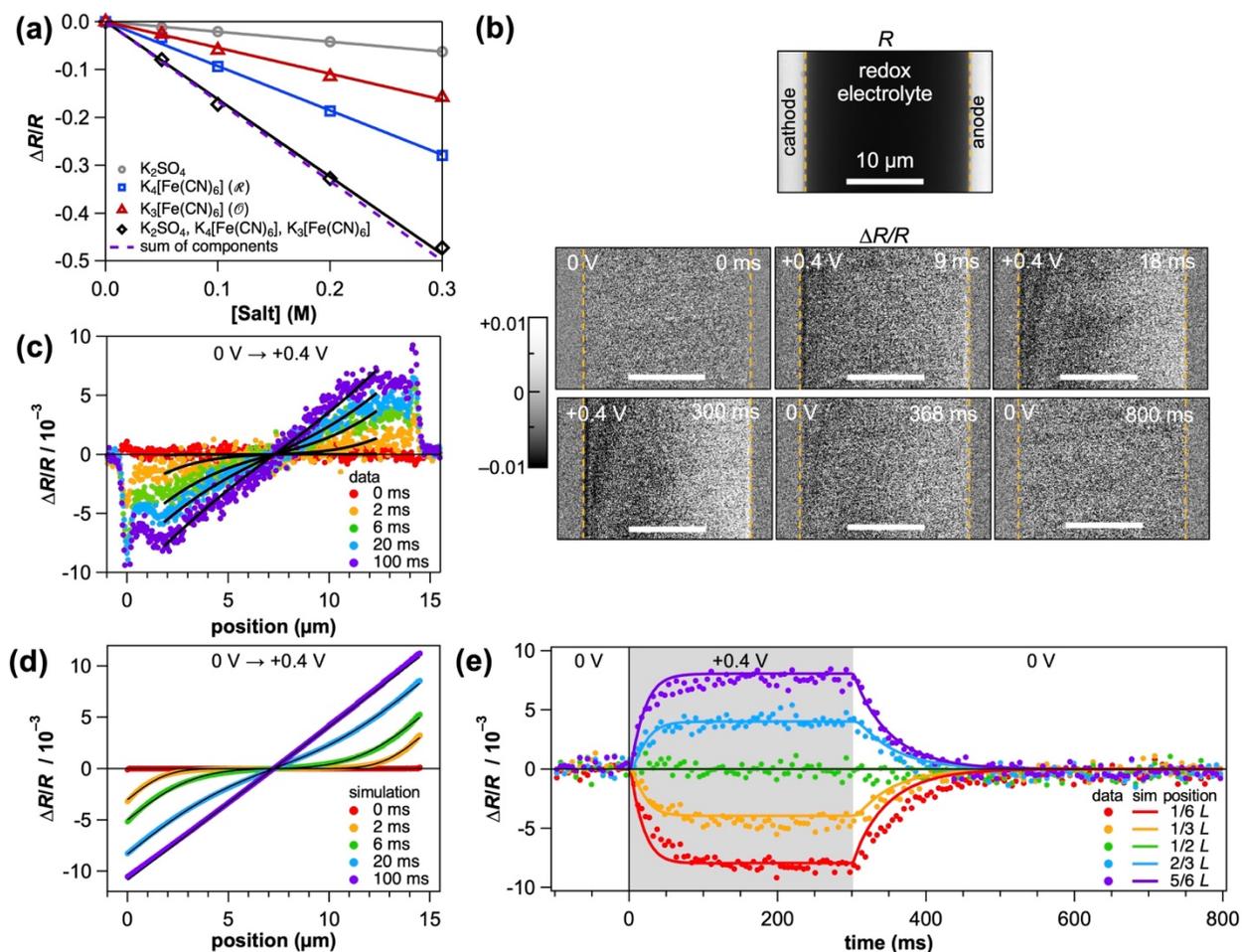

**Figure 2.** Optical detection of lateral ion transport. (a) Static measurement of change in reflectance with respect to solute concentration relative to neat water. (b) Raw reflectance image of microfluidic channel loaded with 100 mM $K_3[Fe(CN)_6]$, $K_4[Fe(CN)_6]$ and $K_2SO_4$, and $\Delta R/R$ time series before (top left), during (remaining top and bottom left frames) and after (bottom right) applying voltage of +0.4 V across the cell. A voltage of +0.4 V was applied from $t = 0$ to 300 ms and was otherwise at ground. Right-hand electrode is the positive terminal (anode). Electrode edges are indicated with dashed yellow lines. (c) Profile of $\Delta R/R$ versus position as a function of time after applying +0.4 V at $t = 0$. Traces were averaged over 20 μm along the channel axis (vertical as pictured in (b)) and over 2 ms in time. Black curves show fits to eq 3. (d) Simulated $\Delta R/R$ profiles that parallel the experimental data in (c). Black curves show fits to eq 3. (e) Experimental (circles) and simulated (curves) time traces of $\Delta R/R$ at a series of 6 different positions between electrodes relative to the channel width $L$ (horizontal as pictured in (b)). Experimental time traces were averaged 0.5 μm across the cell width and over 20 μm along channel axis.

To validate the interpretation that $\Delta R/R$ directly reflects ion gradient evolution in solution, we performed finite element simulations of the electrochemistry and mass transport. The one-dimensional model consisted of two parallel facing Pt electrodes (14.5 μm gap) in a 100 mM $K_3[Fe(CN)_6]$, $K_4[Fe(CN)_6]$, and $K_2SO_4$ electrolyte solution (see SI for details). Mimicking the conditions of the experiments, all species in solution began at uniform concentration, then a voltage of +0.4 V was applied for a duration of 300 ms, at which point the voltage was switched to zero. The simulations yield the concentrations of all species as a function of time and position (Figure S3), from which the experimentally observed $\Delta R/R$ was calculated according to Equation 3 (Figures 2d,e, S4). We tuned the exchange current density as the only fit parameter to obtain agreement between the experimental and simulated amplitudes of $\Delta R/R$ at steady state (see SI for details). There may be minor inaccuracies in the optical parameters used in Equation 3 or in the simulation parameters because the model neglects the oxidation of Fe(II) with $O_2$ and because it uses physical parameters, such as the exchange current density and diffusion coefficients, that are only known approximately. Importantly, plots of the calculated $\Delta R/R$ reproduce the essential behavior seen experimentally. $\Delta R/R$ increases near the positive electrode and decreases near the negative electrode (Figure 2d)—approximately reaching steady state around 100 ms—and the system relaxes back to equilibrium after the voltage is turned off (Figure 2e, curves). The small discrepancies in the time traces at different positions across the width of the cell (primarily in the time constants associated with the red and orange locations on the left side of the channel) may be due to asymmetries in the cell that are not accounted for in the model. Thus, the simulations support the attribution of the origin of the $\Delta R/R$ signal as concentration-induced changes in the refractive index.

Because the spatiotemporal data directly reflects ion concentration gradient evolution, it provides quantitative information about the underlying ion transport behavior without relying on models *a priori*. Specifically, we obtain good fits (black curves) of the profiles in Figure 2c,d as a function of position across the cell $x$ and time $t$ using the empirical function,

$$\Delta R/R(x,t) = A(t)(\text{erf}[x/w(t)] - \text{erf}[(L-x)/w(t)]), \text{ (eq 4)}$$

for fixed $L$. This ansatz reflects one-dimensional diffusion of electrochemical reaction products away from each of the electrodes and the phenomenology of the Cottrell equation (Figure 3a).[5,47] The fit range in the experimental data is smaller than in the simulation data in order to avoid electrode-imaging effects (Figure 2c,d). In analogy to empirically obtaining the charge and thermal carrier mean-squared expansion (MSE) in semiconducting or conducting solids by fitting spatiotemporal transient optical microscopy data to a Gaussian of expanding width,[48–52] the fit parameter $w$ in each error function here is a measure of the width of the ion profile being electrochemically generated and then expanding from the surface of each electrode (Figure 3a). We thus calculate the MSE as the temporal change in this width squared, $w^2(t)$, with $w^2(0) \simeq 0$ (Figure 3b, black circles). The experimentally obtained MSE appears linear over the first ~35 ms, suggesting diffusive behavior, and then flattens out, as one would expect when approaching steady state for diffusion in a one-dimensional confined region, corresponding to the timescale on which the ionic species travel across the width of the channel. Indeed, Equation 4 satisfies the diffusion equation in the case that $w^2(t) = 4Dt$, where the factor of 4 is expected for one-dimensional diffusion based on the error function definition. Fitting the MSE to this relationship (yellow curve) therefore gives $D = (7.2 \pm 0.3) \times 10^{-6}$ cm$^2$/s. Performing a similar analysis on the simulation data yields a very similar MSE (cyan curve). The small discrepancy between the experimental and simulated MSE curves could be due to slight mechanical drift in the microscope, the shot noise limit of the CMOS camera during *operando* measurement, assuming dilute solution theory and pure water transport in the simulation, or inaccuracies in the ion diffusivities from the literature used as parameters in the simulations. Also encouragingly, a very similar diffusivity is obtained by fitting the relaxation curves after the voltage is switched off to a diffusion model (Figure S7). The experimentally obtained diffusivity resembles the literature value for [Fe(CN)$_6$]$^{3-}$ used as an input for the simulation (Table S2).[6] This finding suggests that the effective diffusivity measured for such a multicomponent system is dominated by the slowest charged species, analogous to the constraint of charge neutrality on charge carrier transport in semiconductors where the electrostatic attraction of the slowest species 'slows' down the others.[6,53]

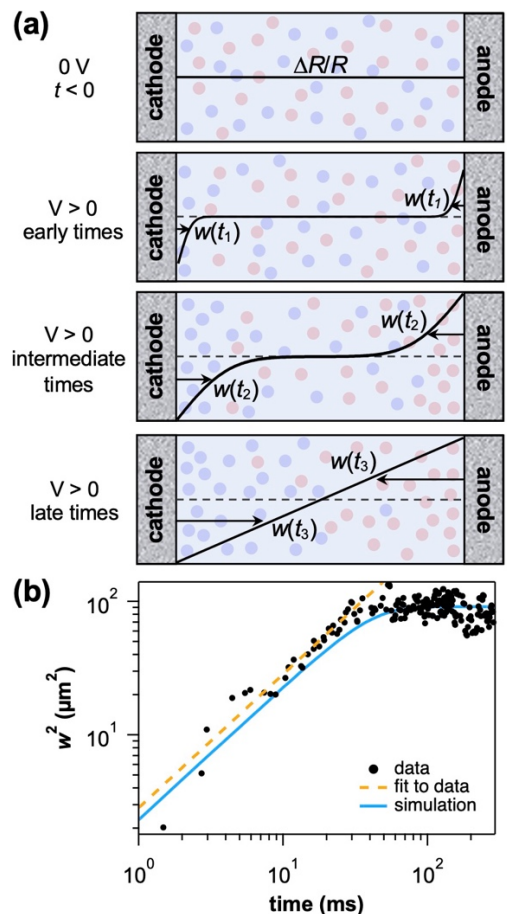

**Figure 3**. Model-free extraction of ion transport. (a) Schematic depiction of the mean-squared expansion of the ion concentration profile, characterized by the distribution's width $w$ as it expands from each electrode. (b) Mean-squared expansion (MSE) of ions from the electrode surfaces calculated from data (black circles) and simulations (cyan curve). Data are fit to $w^2 = 4Dt$ (yellow dashed curve).

We have established that detecting solute concentration-induced differences in index of refraction with high sensitivity enables direct optical imaging of solution-phase ion distributions in the electrolyte and their spatiotemporal evolution. This capability not only allows for quantitative, model-free extraction of ion transport parameters from experimental data but also allows imaging of dynamic responses to electrochemical perturbation with sub-mM sensitivity at millisecond and sub-micrometer resolution. In particular, we have benchmarked this approach using a scenario in which an applied perturbation performs chemistry to alter a solution's makeup while maintaining the ability to determine intrinsic ion diffusivity. Directly probing solution-phase ion transport with high spatial and temporal resolution in native environments without added labels or invasive probes is crucial to understanding the myriad phenomena that rely on ion transport.

We envision a series of far-reaching implications of the methodology developed and validated herein. First, the methodology set forth uses a highly general label-free contrast mechanism that should enable spatiotemporal detection of solutes *and* their transport, as virtually all solutes—including monatomic ions and non-ionic species—perturb at least the real part of the refractive index of their surroundings.[42,43] Even though the characteristic signal strengths that we observe in solution are ~100x smaller than those of ions observed in solid matrices,[41,54] with careful calibration and a sensitive camera, one could envision a complete chemical analysis of electrochemical products being generated *in situ* at an electrode surface. Even the sensitivity to simple salts such as $K_2SO_4$ should be in the mM range in the present configuration (Figure 2a, Table S1). The sensitivity of the method is not so much dependent on the optical response to different ions, which scales with $K_i$ as it is to the net sum of ion contributions as a function space and time; for example, $[Fe(CN)_6]^{3-}$ and $[Fe(CN)_6]^{4-}$ each relatively strongly change the index of refraction of a solution, but their contributions are nearly equal and opposite, and the detection presented in Figure 2 requires sensitivity to this small difference, as exemplified by our measurement-fed model of the individual contributions of the different solution components (Figure S4). This sensitivity is not intrinsic to the method and could conceivably be enhanced by optimizing factors such as the substrate and solution refractive indices or the detector. In the present sample, we estimate that the method is sensitive to probe current densities down to ~10 mA/cm$^2$ on the shortest length and time scales and down to ~$10^{-1}$ mA/cm$^2$ at long times, but the overall range of sensitivity could be extended almost arbitrarily by adjusting the field-of-view, electrochemical cell size and frame rate (see SI for further discussion). Moreover, this strategy is highly accessible because not only is the contrast mechanism general but only modest alterations to a standard optical microscope are required to image electrolyte distributions. As such, we expect this approach will be a valuable label-free probe for ionic transport in a wide range of physical, chemical, and biological contexts, including product formation and collection in (photo)electrochemical energy conversion,[1,3] water splitting,[39] phototriggered ion transport,[55] ion transport in soft matter,[2] aqueous battery function, dissolution dynamics after proper calibration of the refractive index near saturation, microfluidic ion flow, bioelectronic device function, or even label-free electrophysiology dynamics.[4] It could also be readily combined with established strategies to optically detect oxygen and hydrogen evolution,[39] especially since the sensitivity required to see gaseous products is orders of magnitude lesser than what we have demonstrated for solution-phase species, and to potentially probe three-dimensional ion transport in such contexts.[49,56] Second, the MSE analysis used here could serve as a general approach to reveal non-diffusive behavior or spatially-dependent transport parameters. Specifically, similar to the case for carrier transport probed by spatiotemporal methods,[51,52] deviations from diffusive transport manifest as non-linear MSE curves that would be readily discerned. Finally, directly visualizing ion distributions and mass transport spatiotemporally in two dimensions grants information about the impact of electrode structure or heterogeneity and the commensurately local electrode activity that is not accessible from conventional current–voltage measurements that average over the system. As a simple example, variations in the present electrochemical cell's electrode deposition reveal regions of high local activity (Figure S8). Even though the total current integrated over the anode and cathode must be balanced, this constraint is broken locally when an electrode on one side of the channel has a larger local surface area than the one opposite it (Figure S8). The ability to resolve such variations should facilitate electrochemical device design and optimization.

**Notes**
The authors declare no competing financial interest.

**Acknowledgments**
We are grateful to Frédéric Kanoufi, Gerd Rosenblatt, Francis Houle, Steve Harris, Milan Delor, Hannah Weaver and other Ginsberg Group members for helpful discussions. The custom microfluidic electrochemical cell was fabricated by Microfluidics Foundry based in Berkeley California. Instrument development was supported by STROBE, a National Science Foundation Science and Technology Center under grant no. DMR 1548924, and the application to electrochemistry was supported by Liquid Sunlight Alliance, which is supported by the U.S. Department of Energy, Office of Science, Office of Basic Energy Sciences, Fuels from Sunlight Hub under Award Number DE-SC0021266. J.K.U. acknowledges support from the Arnold O. Beckman Postdoctoral Fellowship in Chemical Sciences from the Arnold and Mabel Beckman Foundation. A.J.K. acknowledges funding from the National Science Foundation Graduate Research Fellowship under Grant No. DGE 2146752. L.M.H. acknowledges a National Defense Science and Engineering Graduate Fellowship. N.S.G. acknowledges a David and Lucile Packard Foundation Fellowship for Science and Engineering and a Camille and Henry Dreyfus Teacher-Scholar Award.

**Supporting Materials**
Experimental details, optical spectra, refractive index versus concentration coefficients, simulation details, simulated concentration profiles, component contributions to differential signal, chronoamperometry measurement and simulation, alternating voltage experiments, diffusion modeling, and examples of heterogeneous behavior within electrochemical cell.

# SUPPORTING INFORMATION
# for

# *Operando* Label-free Optical Imaging of Solution-Phase Ion Transport and Electrochemistry

James K. Utterback,[1,2,†,*] Alex J. King,[2,3,4] Livia Belman-Wells,[2,5] David M. Larson,[2,‡] Leo M. Hamerlynck,[1] Adam Z. Weber,[2,3] Naomi S. Ginsberg[1,2,5,6,7,8,*]

[1] Department of Chemistry, University of California, Berkeley, California 94720, United States
[2] Liquid-Sunlight Alliance, Lawrence Berkeley National Laboratory, Berkeley, California 94720, United States
[3] Energy Conversion Group, Lawrence Berkeley National Laboratory, Berkeley, California 94720, United States
[4] Department of Chemical and Biomolecular Engineering, University of California, Berkeley, California 94720, United States
[5] Department of Physics, University of California Berkeley, Berkeley, California 94720, United States
[6] STROBE, National Science Foundation Science and Technology Center, University of California Berkeley, Berkeley, California 94720, United States
[7] Materials Science Division and Molecular Biophysics and Integrated Bioimaging Division, Lawrence Berkeley National Laboratory, Berkeley, California 94720, United States
[8] Kavli Energy NanoSciences Institute at Berkeley, Berkeley, California 94720, United States
[†] Present Address: Sorbonne Université, CNRS, Institut des NanoSciences de Paris, INSP, 75005 Paris, France
[‡] Present Address: Twelve, Berkeley, California 94710, United States

*Correspondence to: nsginsberg@berkeley.edu, james.utterback@insp.jussieu.fr

# Methods

***Interference Reflection Microscopy.*** For all data shown in the text, the light source was a 595 nm light emitting diode (LED) (M595L4, ThorLabs) that was equipped with a $f$ = 32 mm collimating lens (ACL50832U-A, ThorLabs) and driven by a LED driver (T-Cube, ThorLabs). A $f$ = 300 mm wide-field lens focused the beam on the back focal plane of the objective, resulting in wide-field illumination at the sample. A 50/50 beamsplitter reflects the light into a high numerical aperture (1.4 NA) oil-immersion objective (Leica HC PL APO 63×/1.40NA) and onto the sample. Light reflected from the sample–substrate interface was collected through the same objective. The light transmitted through the beamsplitter was focused onto a 12-bit charged metal-oxide semiconductor (CMOS) detector (PixeLINK PL-D752, equipped with the Sony IMX 174 global shutter sensor) using a $f$ = 500 mm lens placed one tube length (200 mm) away from the back focal plane of the objective. The total magnification was 63×500/200 = 157.5. On square pixels that are 5.86 μm in size, this magnification corresponds to 37.2 nm/pixel. Typical region of interest (ROI) sizes range from 1000 × 200 pixels (~37 × 7 μm) to 1000 × 400 pixels (~37 × 15 μm). The image acquisition rate (image exposure time) was determined by the ROI and ranged from 675 Hz (1.5 ms) to 367 Hz (2.7 ms). Data acquisition was implemented in LabVIEW 2022 64-bit. All experiments were performed at 22 °C.

***Sample Preparation.*** Solutions were prepared by dissolving chemicals as-received into ultrapure water (18.2 MΩ·cm). Solutions for *operando* measurements contained 100 mM potassium ferricyanide(III) ($K_3[Fe(CN)_6]$, 99%, Sigma Aldrich), 100 mM potassium hexacyanoferrate(II) trihydrate ($K_4[Fe(CN)_6] \cdot 3H_2O$, >99.5%, Sigma Aldrich) and 100 mM potassium sulfate ($K_2SO_4$, >99%, J.T. Baker). Solutions were prepared in air.

***Spectroscopy.*** UV-Visible absorption measurements were conducted on a Cary UV-Vis 100 spectrophotometer (Agilent, USA). Measurements were performed at 2.5 mM $K_3[Fe(CN)_6]$ or $K_4[Fe(CN)_6]$. The spectrum of the LED used as a light source in reflectance measurements was collected using an Ocean Optics USB4000, measured at the sample location after the objective and after filtering with 505 nm long-pass and 650 nm short-pass filters.

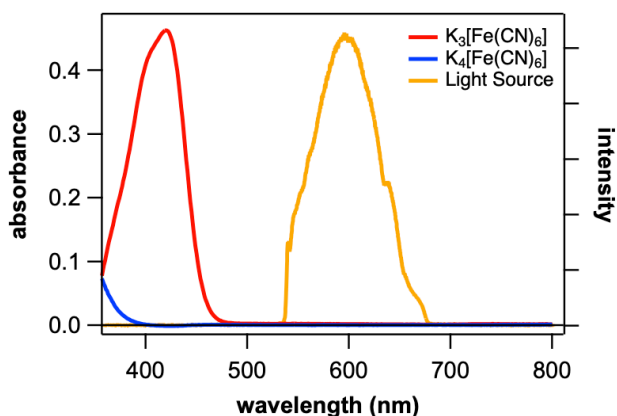

**Figure S1**. UV-visible absorption spectrum of redox electrolyte components and light source spectrum at sample. Absorption spectra are for 2.5 mM solutions in water in a 1 cm-pathlength cuvette. Both the reduced and oxidized states of the iron complex exhibit negligible absorbance in the region that overlaps the light source spectrum, indicating a negligible imaginary part of the refractive index.

***Static Reflectance vs. Concentration Measurements.*** The change in reflectance with respect to solute concentration for all relevant species (Figure 2a) was measured using droplets of solution on a blank coverslip. A series of solutions with known concentrations were prepared, and one at a time a large droplet of each solution was placed on a coverslip. Enough volume was used such that the distance from the coverslip surface to the top of the droplet (~1 mm) is much greater than the depth of field (<1 µm), ensuring that only light reflected from coverslip–solution interface arrives at the camera. We focused on the coverslip–solution interface by imaging the reflection of a focused beam, then the reflected light intensity of the widefield 595 nm LED beam was recorded. The light intensity on the camera $I$ is proportional to $R$, thus the change in the intensity relative to neat water due to solvated species is equal to the change in reflectance $\Delta R/R = \Delta I/I$.

The results of $\Delta R/R$ *versus* concentration for $K_3[Fe(CN)_6]$, $K_4[Fe(CN)_6]$, $K_2SO_4$ and their combination in equal parts appear in Figure 2a. Fits of the data to Equation 3 in the main text give the values of $K$ for each salt compound, which are summarized in Table S1.

**Table S1**

| Solute | $K$ |
|---|---|
| $K_2SO_4$ | $1.0 \pm 0.1$ |
| $K_3[Fe(CN)_6]$ (Fe(III)) | $2.6 \pm 0.2$ |
| $K_4[Fe(CN)_6]$ (Fe(II)) | $4.4 \pm 0.2$ |
| 1:1:1 $K_2SO_4$:$K_3[Fe(CN)_6]$:$K_4[Fe(CN)_6]$ | $7.8 \pm 0.3$ |

***Electrochemical Cell Fabrication and Use.*** The electrochemical cell used for all experiments in the text was custom fabricated (Microfluidic Foundry, Berkeley, CA) to allow for imaging lateral ion transport between vertical electrodes by interference reflection microscopy. The fabrication is based on deep-reactive ion etching (DRIE), lift-off, and PDMS–quartz bonding. First, a 3 µm deep reaction channel, together with other fluidic components, is DRIE etched on a 180 µm-thick quartz wafer. Then metal layers of 10 nm-thick chromium and 200 nm-thick platinum are sputter deposited and patterned by lift-off process. These metal patterns form the electrodes and the connections to peripheral instruments. The 10 nm chromium layer is the adhesion layer between the platinum and the quartz surface while platinum acts as the electrochemically-active electrode. After both the fluidic and the electrical components are patterned, the wafer is diced into 24 × 50 mm devices. The device is then capped by a PDMS cover by plasma-aided PDMS–quartz bonding. The PDMS layer was prepared using SYLGARD™ 184 Silicone Elastomer. The access holes for both fluidic sample loading and electrical connections are punched on the PDMS cover. Thin, low-stiffness wires are attached to the square metal pads outside the PDMS using silver epoxy in order to connect to the potentiostat.

The nominal surface area of each electrode is $6 \times 10^{-5}$ cm$^2$, but the true active surface area is likely different from this due to imperfections in fabrication.

Solution is introduced to the electrochemical cell using two syringe pumps and syringes connected to the PDMS inlet/outlet via pipette tips—one syringe applies negative pressure to the outlet while the syringe loaded with solution applies positive pressure to the inlet, each at ≤1 atm. During experiments, the inlet and outlet are plugged to prevent evaporation. The total volume inside the channel, including the reservoirs formed by the PDMS holes, is ~100 nL.

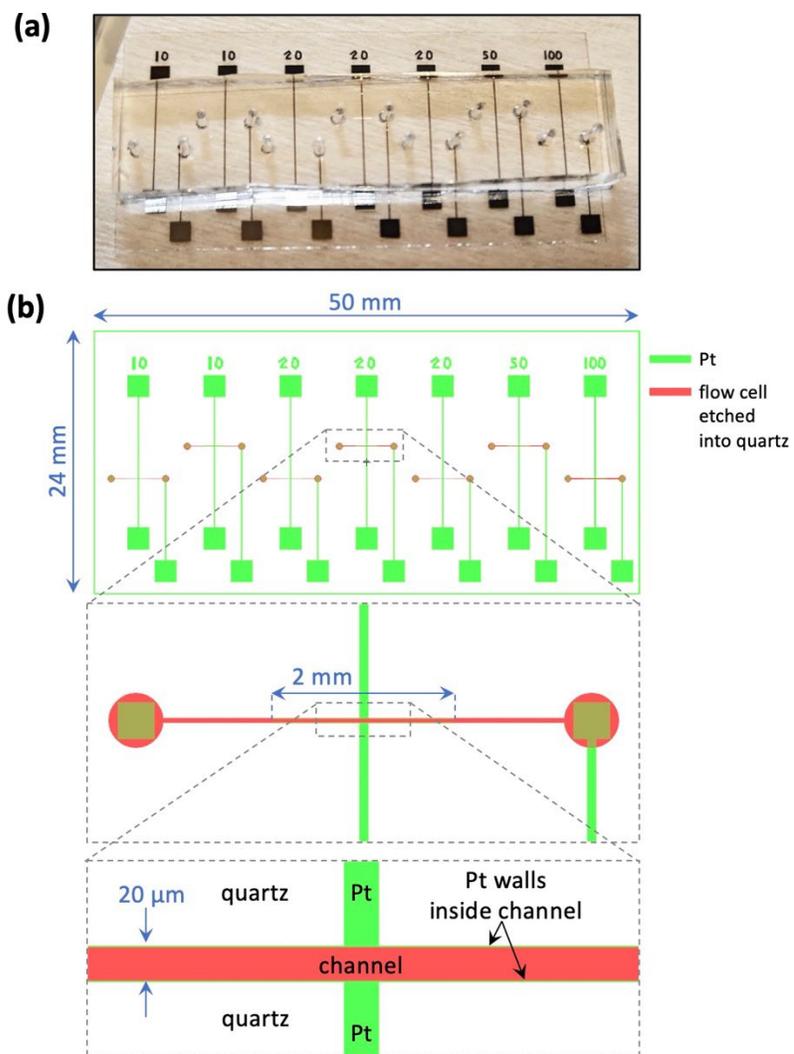

**Figure S2**. Pattern for electrochemical cell design. The numbers above the electrode indicate the nominal width of the corresponding channel, though only the "20 μm" channel was used in this work.

***COMSOL Simulations.*** Finite element simulations of electrochemical system were performed using COMSOL Multiphysics® 6.0 (COMSOL Inc., Los Angeles, CA). Simulations were conducted using the Electrochemistry Module. The PARDISO general solver was used with a general tolerance of 0.001 and a physics-controlled mesh of 545 elements. Steady-state species conservation governs the electrochemical transport for $[Fe(CN)_6]^{4-}$, $[Fe(CN)_6]^{3-}$, $OH^-$, $H^+$, and $K^+$ throughout the one-dimensional system,[1]

$$\nabla \cdot N_i = R_{B,i}, \quad (S1)$$

where $N_i$ is the flux of species $i$, and $R_{B,i}$ is a volumetric source term describing generation of species $i$ due to homogeneous buffer reactions. The Nernst-Planck equation is used to calculate the molar flux of species $i$,[1]

$$N_i = -D_i \nabla c_i - \frac{z_i F}{RT} D_i c_i \nabla \phi_L, \quad (S2)$$

where $D_i$, $c_i$, $z_i$ are the diffusivity, concentration, and charge of species $i$, respectively, $\phi_L$ is the liquid-phase potential of the electrolyte, $F$ is Faraday's constant, $R$ is the ideal-gas constant, and $T$ is the temperature. The first term captures the transport of species by diffusion, and the second term describes charged species migration. The diffusivities are provided in Table S2. To solve for the liquid-phase potential, electroneutrality is enforced,

$$\sum_i z_i c_i = 0. \quad (S3)$$

**Table S2.** List of model parameters for 25 °C and their sources

| Parameter | Value | Units | Ref. |
|---|---|---|---|
| $D_{H^+}$ | $9.31 \times 10^{-5}$ | $cm^2\ s^{-1}$ | 1 |
| $D_{OH^-}$ | $5.26 \times 10^{-5}$ | $cm^2\ s^{-1}$ | 1 |
| $D_{K^+}$ | $2.05 \times 10^{-5}$ | $cm^2\ s^{-1}$ | 1 |
| $D_{[Fe(CN)_6]^{3-}}$ | $7.39 \times 10^{-6}$ | $cm^2\ s^{-1}$ | 1 |
| $D_{[Fe(CN)_6]^{4-}}$ | $8.96 \times 10^{-6}$ | $cm^2\ s^{-1}$ | 1 |
| $K_w$ | $1 \times 10^{-14}$ | $mol\ L^{-1}$ | 2 |
| $k_w$ | $1.6 \times 10^{-3}$ | $mol\ L^{-1}\ s^{-1}$ | 2 |
| $i_{0,1}$ | 17 | $mA\ cm^{-2}$ | fitted |
| $i_{0,2}$ | $1 \times 10^{-3}$ | $mA\ cm^{-2}$ | 4 |
| $i_{0,3}$ | 17 | $mA\ cm^{-2}$ | fitted |
| $\alpha_1$ | 0.5 | — | 3 |
| $\alpha_2$ | 0.5 | — | 4 |
| $\alpha_3$ | 0.5 | — | 3 |

The only homogeneous bulk reaction that occurs is water dissociation, which is captured by the source term, $R_{B,i}$, and is

$$H_2O \xrightleftharpoons[k_{-w}]{k_w} H^+ + OH^-, K_w \quad (S4)$$

where $k_{n/-n}$ and $K_n$ are the rate constants and equilibrium constant for reaction $n$, respectively. These constants are provided in Table S2. $k_{-n}$ is calculated by

$$k_{-n} = \frac{k_n}{K_n}, \tag{S5}$$

and $R_{B,i}$ is given by[2]

$$R_{B,i} = \sum_n s_{i,n} c_{ref} \left( k_n \prod_{s_{i,n}<0} a_i^{-s_{i,n}} - \frac{k_n}{K_n} \prod_{s_{i,n}>0} a_i^{s_{i,n}} \right), \tag{S6}$$

where $s_{i,n}$ is the stoichiometric coefficient for species $i$ in reaction $n$, and $a_i$ is the activity of species $i$. $c_{ref}$ is a reference concentration defined as 1 M.

The boundary conditions at the electrode surfaces are Neumann boundary conditions specified by Faraday's law and the concentration-dependent Tafel equation,

$$i_k = -i_{0,k} \left( \frac{c_k}{c_{ref}} \right) \exp\left( -\frac{\alpha_k F}{RT} (\phi_s - \phi_L - U_{0,k}) \right), \tag{S7}$$

$$N_k = -\frac{v_k i_k}{nF}, \tag{S8}$$

where $\phi_s$ is the solid phase electrode potential, $\phi_L$ is the liquid phase electrolyte potential, and $c_k$, $i_{0,k}$, $\alpha_k$, $U_{0,k}$, and $v_k$ are reactant concentration, the exchange current density, transfer coefficient, equilibrium potential, and reaction stoichiometric coefficient for product $k$, respectively. Moreover, $n$ is the number of electrons transferred in the electrochemical reaction. At the cathode surface, the following electrochemical reactions occur:[5]

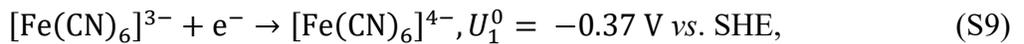
$$[Fe(CN)_6]^{3-} + e^- \rightarrow [Fe(CN)_6]^{4-}, U_1^0 = -0.37 \text{ V } vs. \text{ SHE}, \tag{S9}$$

$$2H_2O + 2e^- \rightarrow H_2 + 2OH^-, U_2^0 = 0 \text{ V } vs. \text{ SHE}. \tag{S10}$$

When accounting for the Nernstian pH shift (pH = 7), the equilibrium potential of the water reduction reaction becomes −0.407 V vs. SHE (standard hydrogen electrode). In the experiments we thus chose to apply a potential of 0.4 V in order to avoid significant contributions from water reduction, though a small amount of water reduction could occur at the applied potential.

At the anode surface, the following reaction occurs:

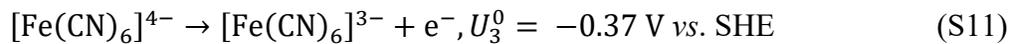
$$[Fe(CN)_6]^{4-} \rightarrow [Fe(CN)_6]^{3-} + e^-, U_3^0 = -0.37 \text{ V } vs. \text{ SHE} \tag{S11}$$

At the center of the system ($L/2$), the species concentrations are set to their bulk values (0.1 M $[Fe(CN)_6]^{4-}$ and $[Fe(CN)_6]^{3-}$ solution) in order to lock the solution at a specific point in the modeling domain; otherwise, infinitely many solutions are possible. This constraint is physical given the symmetry of the cell.

The resulting simulated concentration profiles after applying +0.4 V at time $t = 0$ appear in Figure S3. The simulated $\Delta R/R$ profiles over time and time traces at selected positions across the cell in Figure 2 of the main text are calculated from the concentration profiles of Figure S3 by using Equation 3. Figure S4 demonstrates how, at steady state as an example, the calculated contributions from $[Fe(CN)_6]^{4-}$ and $[Fe(CN)_6]^{3-}$ partially cancel to give the total $\Delta R/R$ profiles. The resulting simulated total current density transient appears in Figure S5.

The exchange current densities at the anode and cathode, $i_{0,1}$ and $i_{0,3}$, were used as a common fit parameter in the simulations. Their values were held equal to each other and varied until the resulting maximum amplitude of $\Delta R/R$ at steady state (>100 ms) matched the experimental value (Figure 2e). The final value of 17 mA/cm$^2$ (Table S2) is in good agreement with the expected range for facile reactions on Pt.[1] Uncertainties in other simulation parameters, optical parameters, or the approximate form of Equation 3 in the manuscript could also be at play, but exchange current density is expected to have the largest uncertainty in this system and the success of using a single fit parameter suggests that the model captures the essential behavior of the system.

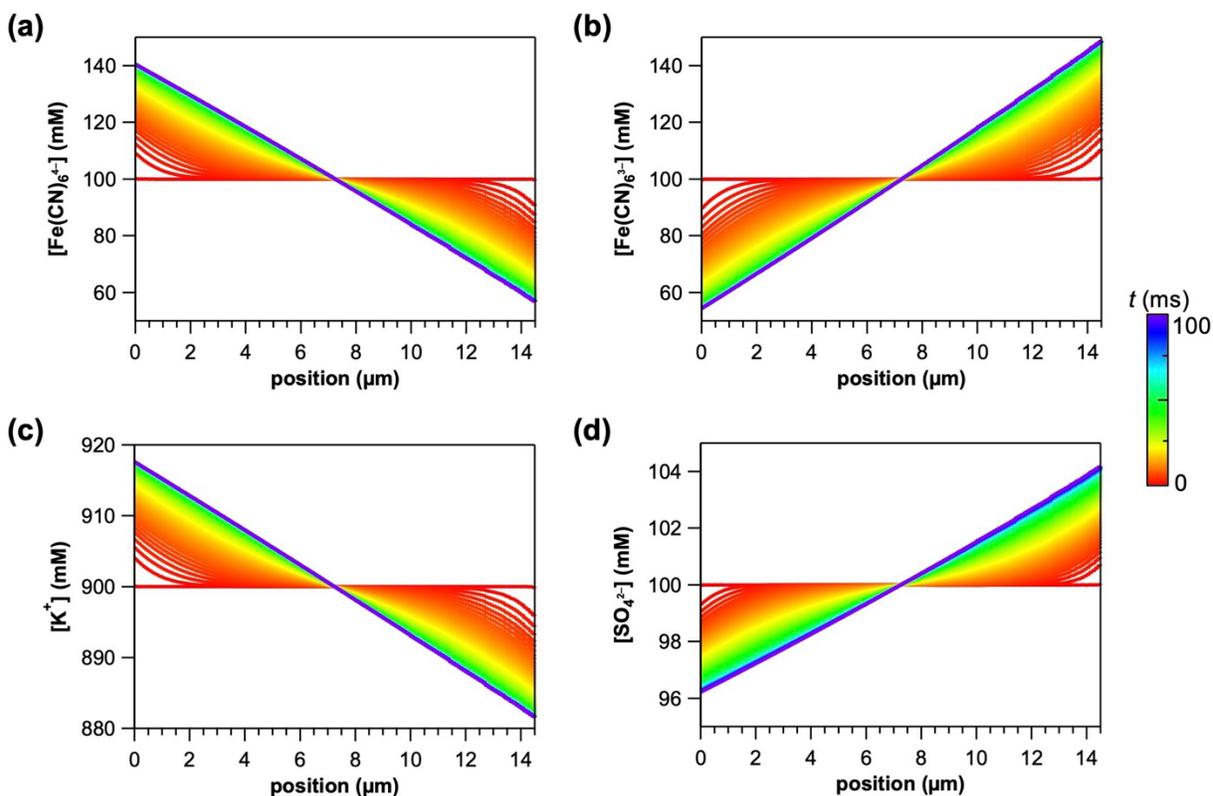

**Figure S3**. Simulated concentration profiles of (a) $[Fe(CN)_6]^{4-}$, (b) $[Fe(CN)_6]^{3-}$, (c) K$^+$, and (d) SO$_4^{2-}$ across the channel after a voltage of +0.4 V is applied at time $t = 0$.

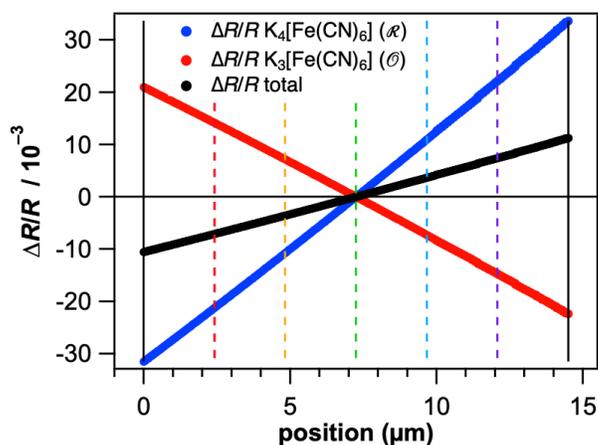

**Figure S4.** Calculated steady-state contributions of $K_4[Fe(CN)_6]$ (blue dots) and $K_3[Fe(CN)_6]$ (red dots) to $\Delta R/R$ and their sum (black dots) at 300 ms after application of +0.4 V at time $t = 0$. Vertical dashed lines correspond to locations of time traces in Figure 2e.

*Current–Voltage Measurements.* Current–voltage measurements were achieved with a Keithley 2450 SourceMeter. Voltage, current and timestamps were collected at ~745 Hz with fixed voltage and current ranges, Number of Power Line Cycles (NPLC) set to 0.01 and autozeroing off.

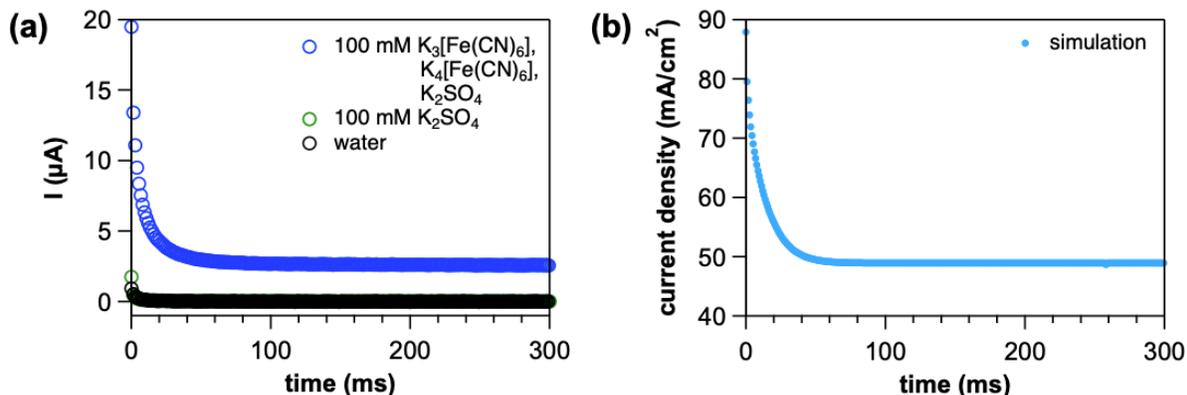

**Figure S5**. (a) Current transient corresponding to data from Figures 2 and 3 in the main text for a solution composed of 100 mM $K_3[Fe(CN)_6]$, $K_4[Fe(CN)_6]$ and $K_2SO_4$. Control experiments, in which the cell was loaded with either 100 mM $K_2SO_4$ or pure water, are also shown. In all cases, a voltage pulse of +0.4 V was applied at time $t = 0$. (b) Simulated total current density transient corresponding to the simulation presented in Figures 2, 3 and S3 for 100 mM $K_3[Fe(CN)_6]$, $K_4[Fe(CN)_6]$ and $K_2SO_4$ after applying a voltage of +0.4 V.

The kinetics of the experimental and simulated current transients reflect the $\Delta R/R$ transients shown in Figure 2e and are in reasonable agreement. The steady-state simulated current density of 49 mA/cm$^2$ (Figure S5b) reproduces the measured steady-state current of 2.6 µA (Figure S5a) in the case that an active area of $5.3 \times 10^{-5}$ cm$^2$ is assumed. This area is in reasonable agreement with the predicted active area, given imperfections in the fabrication process and potential inaccuracies in the simulation parameters. Although one might consider comparing this calculation against cyclic voltammetry, it is subject to the same uncertainties in active electrode area and final solution concentrations that are present in the chronoamperometry data.

S8

**Alternating Voltage**

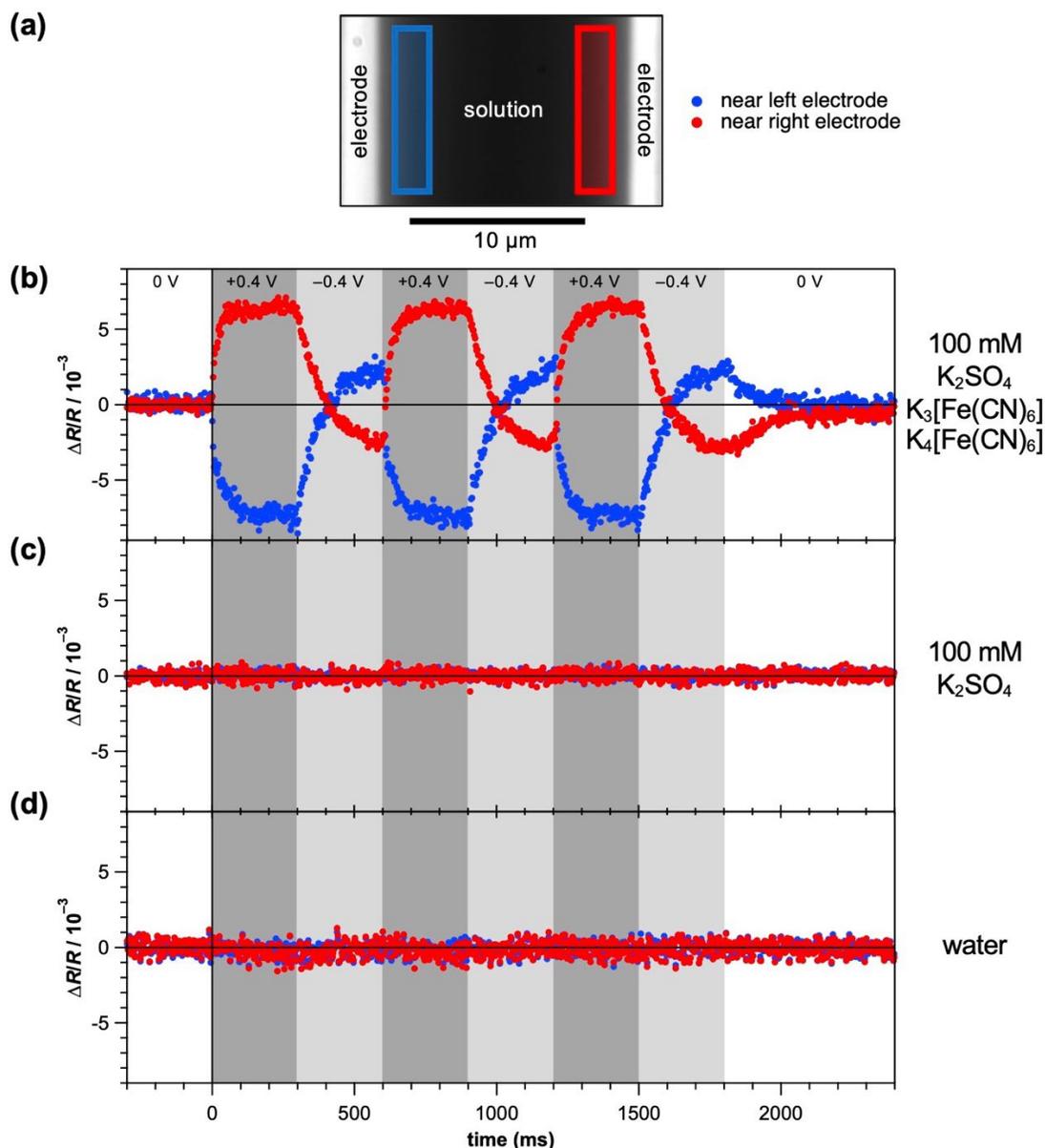

**Figure S6**. Concentration evolution for square wave voltage. The voltage sequence alternated between +0.4 V and –0.4 V three times for 300 ms at each value before ending at no applied voltage and open circuit. (a) Raw reflectance image of the region of the electrochemical cell where measurements were made. The shaded blue and red boxes near the left and right electrodes, respectively, show the regions of interest over which the signal was averaged to obtain the color-corresponding time traces in (b)–(d). $\Delta R/R$ time traces near the left and right electrodes are shown for (b) 100 mM 1:1:1 $K_2SO_4$:$K_3[Fe(CN)_6]$:$K_4[Fe(CN)_6]$, (c) the supporting electrolyte 100 mM $K_2SO_4$ only, and (d) pure water.

## Applying a Diffusion Model to Extract Ion Diffusivity

In addition to extracting the mean squared expansion of solvated ions by the approach detailed in the text, transport parameters can be extracted by alternative means. One straight-forward way to obtain the effective ion diffusion coefficient $D$ of the system from our experiment is to fit the relaxation kinetics once the voltage is turned off. Assuming diffusion-controlled behavior, relaxation from the steady-state concentration gradient back to a uniform equilibrium concentration when the cell is switched from +0.4 V to open circuit should follow the general form of $\Delta R/R(x,t) = \sum_{n=1}^{\infty} \varphi_n(x) \exp(-n^2\pi^2 Dt/L^2)$.[6] According to this form, a short transient that depends on $x$ is followed by exponential asymptotic behavior dominated by the $n = 1$ term with a time constant that is independent of position within the cell. Indeed, such behavior is observed (Figure S7). Fitting the exponential region to $A \cdot \exp(-n^2\pi^2 Dt/L^2)$ gives $D = (7.4 \pm 0.4) \times 10^{-6}$ cm$^2$/s. This diffusivity is consistent with the value found from the mean-squared expansion obtained by fitting the spatial profiles over time (Figure 2c and 3b) as discussed in the manuscript.

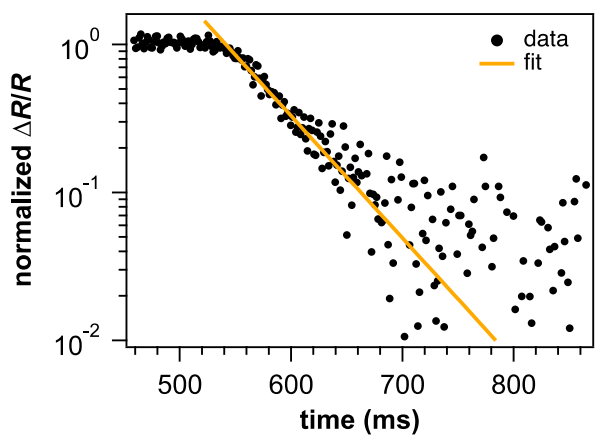

**Figure S7**. Obtaining $D$ assuming a diffusion model corroborates model-free approach in main text.

**Discussion of Sensitivity**

The "current density" in solution is calculated from the well-known expression $J = eD(\Delta\rho/\Delta x)$, which in our case gives ~50 mA/cm$^2$ at steady state. Here, $e$ is the elemental charge, $\Delta\rho$ is a change in ion density, and $\Delta x$ represents some distance along a line perpendicular to the electrodes. To estimate the range of $J$ that can be probed, we consider the limits of the constituent parameters of the expression. The range of $\Delta R/R$, which determines the $\Delta\rho$ that we are sensitive to, is $2^{12}$ on a 12-bit camera. The range of $\Delta x$ is bounded by the diffraction limit (~0.5 μm) on the low end, and an arbitrarily large field of view could be used on the large end. Therefore, overall, the range of measurable current densities is extremely large. In the present system $D = 7 \times 10^{-6}$ cm$^2$/s, such that the lowest detectable $J$ is estimated, using the smallest $\Delta x \sim 0.5$ μm at millisecond time resolution, to be ~10 mA/cm$^2$; at the opposite extreme, using half of the cell ~ 10 μm as the largest length scale of the gradient observable, we obtain a long-time (>100 ms) estimate of $J \sim 10^{-1}$ mA/cm$^2$ sensitivity. Beyond the present system, $D$ can range from infinitely small values (e.g., ions trapped in solids, gels, etc.) up to the fastest relevant species in solution (e.g., H$^+$ in water is $5 \times 10^{-5}$ cm$^2$/s). For example, some extremely fast species could be probed if a large region of interest were used, using a lower-magnification objective; an arbitrarily slow species could be observed by collecting movies at low frame rates over long periods of time, as has been done in the past in the solid-state.[7] More concretely, the present setup should have the resolution to probe H$^+$ in water, provided that a large enough change in concentration were achieved, especially by further optimizing the cell to have a larger separation between electrodes in order to resolve more of the pre-steady-state regime.

Ultimately, the fundamental parameters that govern the experimental sensitivity are the resolutions in space, time, and intensity, which in the present study allow for sub-micrometer and millisecond detection of sub-mM concentration changes.

## Spatiotemporal Heterogeneity

The spatiotemporal method established here also grants direct access to visualizing microscale spatial heterogeneity in the electrode activity and ion flow in two dimensions that is not discernable in commonly-used current–voltage measurements. We highlight some examples in Figure S8.

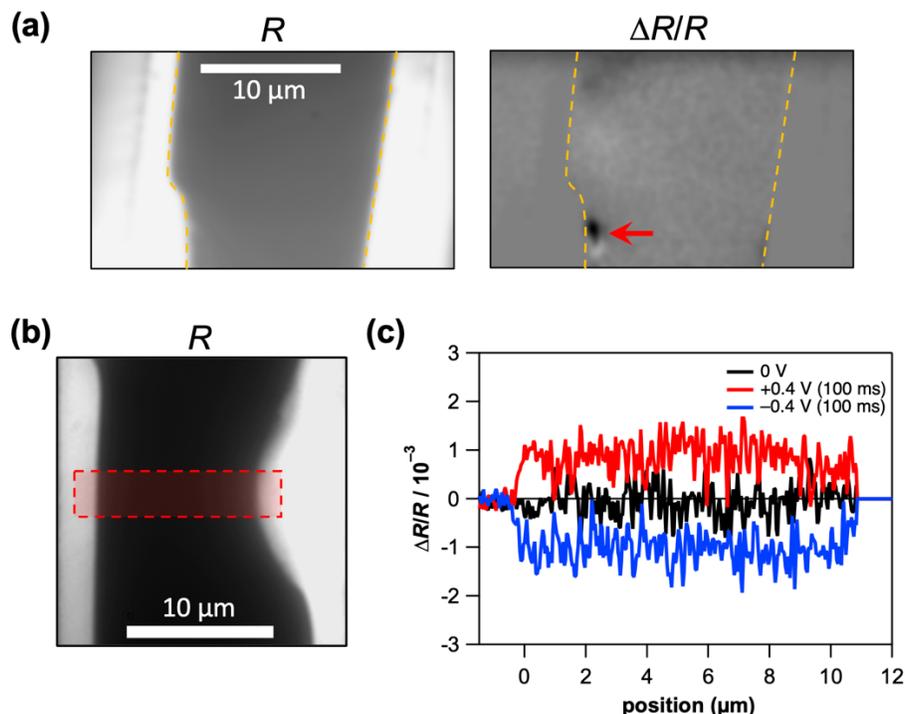

**Figure S8.** Spatiotemporal heterogeneity within the electrochemical cell. (a) High local concentration (right, negative-contrast spot indicated with red arrow) generated near microscale irregularity in electrode shown at left. Electrode edges highlighted with a dashed yellow curve. (b) Image of location in the electrochemical cell where the right-hand electrode extends into the channel with a larger local surface area than the left-hand electrode, generating a local electrochemically-induced concentration profile that deviates substantially from the ideal case. (c) Deviant profile of $\Delta R/R$ versus position across the channel width indicated in (b) after applying the indicated voltage for 100 ms. Traces were averaged over a few µm along channel length (vertical direction as pictured) and over 2 ms in time. Unlike the profiles in Figure 2c that have opposite signs at the cathode and anode, the extent of the one-dimensional $\Delta R/R$ profile across the cell takes on the sign contributed by the larger right-hand electrode in likely because of its larger surface area.